\documentclass[mathleft
]{an}
\usepackage{graphicx}
\usepackage{times}
\begin{document}
\sloppy

\Pagespan{1}{}
\Yearpublication{2007}%
\Yearsubmission{2007}%
\Month{xx}%
\Volume{xxx}%
\Issue{xx}%

\title{Astrophysical supplements to the ASCC-2.5. \\
        Ia. Radial velocities of $\sim$55000 stars and mean radial 
            velocities of 516 Galactic open clusters and associations
        }

\author{N.V. Kharchenko\inst{1,2,3}
\and R.-D. Scholz\inst{1}\fnmsep\thanks{Corresponding author:
  \email{rdscholz@aip.de}\newline}
\and A.E. Piskunov\inst{1,3,4} 
\and S. R\"{o}ser\inst{3} 
\and E. Schilbach\inst{3}
}
\titlerunning{Astrophysical supplements to the ASCC-2.5}
\authorrunning{Kharchenko et al.}
\institute{
Astrophysikalisches Institut Potsdam, An der Sternwarte 16, 
D-14482 Potsdam, Germany
\and 
Main Astronomical Observatory, 27 Akademika Zabolotnogo St.,
03680  Kiev, Ukraine
\and 
Astronomisches Rechen-Institut, M\"{o}nchhofstrasse 12-14, D-69120
Heidelberg, Germany
\and
Institute of Astronomy of the Russian Acad. Sci.,
48 Pyatnitskaya St., 119017 Moscow, Russia
}

\received{25 Jan 2007}
\accepted{23 Apr 2007}
\publonline{...}


\abstract{%
  We present the 2nd version of the Catalogue of Radial Velocities 
  with Astrometric Data (CRVAD-2). This is the result of 
  the cross-identification of stars from the All-Sky Compiled Catalogue of 2.5 
  Million Stars (ASCC-2.5) with the General Catalogue of Radial Velocities 
  and with other recently published radial velocity lists and catalogues. The
  CRVAD-2 includes accurate J2000 equatorial coordinates, proper motions and
  trigonometric parallaxes in the Hipparcos system, $B, V$ photometry in
  the Johnson system, spectral types, radial velocities (RVs), multiplicity 
  and variability flags for 54907 ASCC-2.5 stars.  
  We have used the CRVAD-2 for a new determination of mean RVs 
  of 363 open clusters and stellar associations considering their established
  members from proper motions and photometry in the ASCC-2.5. For 330
  clusters and associations we compiled previously published mean RVs 
  from the literature, critically reviewed and partly revised them. The
  resulting Catalogue of Radial Velocities of Open Clusters and
  Associations (CRVOCA) contains about 460 open clusters and 
  about 60 stellar associations in the Solar neighbourhood.
  These numbers still represent less than 30\% of the total number of
  about 1820 objects currently known in the Galaxy. The mean
  RVs of young clusters are generally better known than those of older ones.
  }

\maketitle

\section{Introduction}

In this paper we present the 2nd version of the 
Catalogue of Radial Velocities (RVs) of
Galactic stars with high precision Astrometric Data (CRVAD-2). As
a first application of this enlarged data set we have compiled
a new Catalogue of Radial Velocities of Open Clusters and
Associations (CRVOCA) currently known in the Galaxy. 

The first version of the CRVAD (Kharchenko, Piskunov \&
Scholz 2004a, hereafter Paper~I) was the result of the cross-identification of
the All-Sky Compiled Catalogue of 2.5 Million Stars (ASCC-2.5;
Kharchenko 2001) with the General Catalogue of Radial Velocities (GCRV;
Barbier-Brossat \& Figon 2000). The CRVAD included about 33\,500
stars. Since some new big RV catalogues
have been published recently, including original observational data 
(Nordstr\"om et al.~2004; Famaey et al.~2005) as well as new RV compilations
(Goncharov 2006), we have built an extended RV list
of about 55\,000 stars which now forms the 2nd CRVAD version.

A new ambitious RV survey of the southern sky, the RAdial Velocity Experiment 
(RAVE), has been initiated a few years ago as a wide international 
collaboration  
(Steinmetz 2003). The first RAVE data release containing RVs for about 25\,000 
stars, has already been published (Steinmetz et al.~2006), the second data release 
is expected in early 2007 and will probably contain about twice as many stars. We 
have not yet included RAVE measurements in the construction of the CRVAD-2 so that 
the latter can be considered as the pre-RAVE status on stellar RVs in the Galaxy.

More than 200\,000 RVs of generally fainter stars are included in the 
latest Sloan Digital Sky Survey (SDSS) data release (DR5; Adelman-McCarthy 
et al. 2007).  This number will be more than doubled by the Sloan Extension 
for Galactic Understanding and Exploration (SEGUE) project targetting stars 
fainter than 14th magnitude in about 200 sky fields covering the sky visible 
from the northern hemisphere (see e.g. Re Fiorentin et al. 2007). The SDSS 
and SEGUE RVs are of lower accuracy ($\sim$10-20~km/s; Ivezic 2006) 
than those of the RAVE survey (typically $\sim$2~km/s; Steinmetz et al.~2006)

The ASCC-2.5 is based on stellar data from the catalogues of the Hipparcos-Tycho 
family, 
including the Tycho-2 catalogue,
and provides the most complete all-sky catalogue of stars having uniform 
high precision astrometric and photometric data down to $V \approx$ 14 with a
completeness limit at $V \approx$ 11.5. The ASCC-2.5 has been the main tool for 
our studies of open clusters. Based on ASCC-2.5 data we identified 513 known 
open clusters and 7 known compact stellar associations (Kharchenko et al.~2005a) 
and detected 130 new open clusters (Kharchenko et al.~2005b). A uniform 
combined spatial-kinematical-photometric cluster membership analysis (Kharchenko 
et al.~2004b) was implemented for all 650 objects and new uniform scales 
of cluster properties (structure, photometry, evolution, and kinematics) were 
established. All these properties were determined from cluster members in the 
ASCC-2.5, and mean cluster radial velocities $\overline{\rm{RV}}$s were 
obtained from cluster members in the CRVAD. In this paper 
we have used the CRVAD-2 for 
determining and/or revising the $\overline{\rm{RV}}$s for all open clusters
and associations with available measurements.  

Distinguishing (young) ''open clusters'' and ''stellar associations'' (or ''OB
associations'') is not a trivial task (Brown 2001),
and one can find different classifications of the same object in the
literature. For instance, the well known open cluster
Melotte 20 ($\alpha$ Per) is also being called the Per OB3 association
(Mel'nik \& Efremov 1995; de Zeeuw et al.~1999),
whereas the Cyg OB2 association is also being studied as a (super-star)
cluster (Hanson 2003) or even as a young globular cluster (Kn\"odlseder 2000) 
based on its mass estimates. 
Our list of 650 objects identified in ASCC-2.5 data therefore includes 
several cluster-like associations (e.g. Vel~OB2, Sco~OB4). On the other hand,
some objects used to be called open clusters and we continue to do so, 
although they could 
be considered as associations according to their ages and stellar content.
Several open clusters represent association cores (e.g. ASCC~16; Brice\~{n}o 
et al.~2005; Kharchenko et al.~2005b). In order to get 
a complete picture on the current status of open cluster radial velocities,
we decided to include not only Galactic open clusters but also the known 
stellar associations in the Galaxy in the construction of the CRVOCA.

For the last decades stars in clusters have been much more attractive targets
for RV measurements than field stars. This is obvious from looking at the sky
distribution of bright stars with available RV measurements (see e.g. Fig.~5 
in Paper~I). The increased number of stars in the CRVAD-2 still represents
only about 2\% of the ASCC-2.5 stars over the whole sky. However, the RV stars 
are concentrated in sky areas with open clusters, where about 5\% of 
the ASCC-2.5 stars have RV measurements. For fainter stars the non-uniform
distribution of RV measurements over the sky is even more pronounced since
many special observing programs are being carried out for fainter open
cluster members in order to obtain cluster $\overline{\rm{RV}}$s.

Nevertheless, our knowledge on open cluster $\overline{\rm{RV}}$s  
is much poorer than on their proper motions. According to Dias et al.~(2002) 
(we refer here to the updated online database, v. 2.7, 
2006/10/27 at http://www.astro.iag.usp.br/\~{}wilton/), $\overline{\rm{RV}}$s 
have been published for only 361 out of about 1760 known clusters, 
All these data are very inhomogeneous: sometimes 
the $\overline{\rm{RV}}$ of a cluster is taken from
measurements of only one star, sometimes the $rms$ errors reach 30~km/s, and
some authors did not give any information on the accuracy at all. 

In building the CRVOCA we also included a critical review of all
available data on $\overline{\rm{RV}}$s of open clusters and associations
in the literature.
In cases where several literature values were available, we gave
preference to the most reliable data. In that respect, we reviewed the number
of cluster members used for the determination of the $\overline{\rm{RV}}$s, 
the applied method of membership determination, and the year of publication. 

In Sec. 2 we describe the construction of the CRVAD-2. 
Sec. 3 includes statistics on the objects in the CRVAD-2 and a summary 
of its content. In Sec. 4  we give the details on the construction of the 
CRVOCA and summarise its content, and in Sec. 5 we consider the new census 
of $\overline{\rm{RV}}$s of open clusters and associations and compare our
own determinations of $\overline{\rm{RV}}$s with data from the literature.

\section{Construction of the CRVAD-2}

The CRVAD-2 is the result of updating and expanding the list of stars with 
known RVs and high precision astrometric and photometric data. These data were
taken from the ASCC-2.5 catalogue (Kharchenko 2001), which is mainly based on 
the Hipparcos and Tycho catalogues, including the Hipparcos Multiple System 
Annex. The ASCC-2.5 contains equatorial coordinates and proper motions in the 
Hipparcos system, as well as $B, V$ stellar magnitudes in the Johnson system. 
Additionally, trigonometric parallaxes, spectral types, multiplicity and 
variability flags (if avaliable in Hipparcos and Tycho-2), as well as HD and 
BD designations are given.    

The RVs in the first version of the CRVAD (Paper~I) were taken from
the GCRV, which supplemented  the Wilson-Evans-Batten catalogue (Dufolt,
Figon \& Meyssonier 1995) with observations published until December 1999. 
In the cross-identification process of the ASCC-2.5 with the GCRV
all the relevant information (including identifiers, coordinates) was used.
A detailed description of the cross-identification was given in
Kharchenko et al (2004b). The first version of the CRVAD contained 34\,553 
stars from the ASCC-2.5 identified with 33\,509 stars from the GCRV.

\begin{figure}[t]
\includegraphics[width=80mm,height=80mm,angle=270]{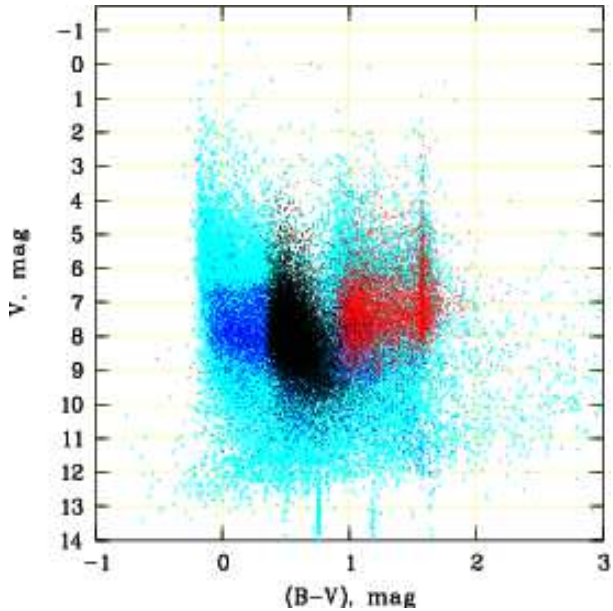}
\caption{Colour-magnitude diagram of the CRVAD-2 stars
(cyan points). Black, red and blue points correspond to stars
from Nordstr\"om et al.~(2004),  Famaey et al.~(2005) and
Goncharov~(2006), respectively.}
\label{fig_v_bv}
\end{figure}



\begin{table}
\caption{Statistics of the CRVAD-2 according to the source catalogues }
\label{tab_soc}
\setlength{\tabcolsep}{2pt}
\begin{tabular}{lcr}
\hline
Source catalogue&Index& Number of \\
                &     & ASCC-2.5 stars\\
\hline
Famaey et al.~(2005)        & 1 & 3488\\
Goncharov (2006)           & 2 & 5756\\
Nordstr\"om et al.~(2004)     & 3 &15245\\
CRVAD (Barbier-Brossat \& Figon 2000)& 4 &27636\\
Average CRVAD \& Famaey et al.~(2005)  & 5 & 2782\\
\hline
\end{tabular}
\end{table}

Three recently published RV catalogues, which we used for expanding our
RV data base into the CRVAD-2, are briefly described in the following:

Nordstr\"om et al.~(2004) published their results of the Geneva-Copenhagen 
survey of the Solar neighbourhood for nearby F and G dwarfs. Their RV 
catalogue includes 14\,139 stars, the majority of which (about 13\,500 stars) were
measured by Nordstr\"om et al.~(2004) with the photoelectric cross-correlation
spectrometers CORAVEL. The RVs for 675 of their stars were taken from the 
GCRV, and therefore we did not consider them as new data. On the other hand,
about 2400 of the stars with newly determined RVs were already included in
the GCRV of Barbier-Brossat \& Figon (2000), where some of these stars were
listed with preliminary observational results. In the construction of the
CRVAD-2 we replaced the data of these stars with the final results from 
Nordstr\"om et al.~(2004). In the cross-identification of the Nordstr\"om 
et al.~(2004) catalogue with the ASCC-2.5 we again used all available data, 
including the coordinates and component identifiers of multiple stars.

Famaey et al.~(2005) published their results of the local kinematics of 6\,691
K and M giants from CORAVEL RVs and Hipparcos/Tycho-2 proper motion data. 
After excluding the binaries for which no center-of-mass RV
could be measured, their catalogue includes 6\,029 Hipparcos stars. 
For about 2\,800 of these stars, RVs were already listed in the GCRV, and 
we computed the weighted means of the two RV values and included them 
in the CRVAD-2. The cross-identification with the ASCC-2.5 was carried out 
by Hipparcos numbers.

Recently Goncharov (2006) presented the Pulkovo Compilation of Radial
Velocities (PCRV), where he included the RVs of about 35\,500 Hipparcos
stars taken from all the above mentioned big catalogues as well as from 
practically all smaller lists with heliocentric RVs published until 2004.
Only the data of the latter lists were taken for completing
the CRVAD-2 list of stars. Here we used again the Hipparcos numbers for the
cross-identification. Note that part of the PCRV stars have also been published 
and analysed as the Orion Spiral Arm CAtalogue (OSACA) by 
Bobylev, Goncharov \& Bajkova (2006).

\section{Statistics and content of the CRVAD-2}

Altogether 54\,907 stars from the ASCC-2.5 were identified with 51\,762 stars
from the RV source catalogues, 3\,085 stars have secondary components and 30 
stars have 3rd components in multiple systems. In cases where an ASCC-2.5 multiple  
star was not resolved in the RV source catalogue, the same RV value was asigned to 
all components. 
These objects were flagged in the CRVAD-2.
Table~\ref{tab_soc} gives the
numbers of stars in the CRVAD-2 according to the source catalogues but
including the components of multiple systems. The index corresponds to the 
designation of a source catalogue in the CRVAD-2.

The apparent colour-magnitude diagram of the CRVAD-2 objects is shown in 
Fig.~\ref{fig_v_bv}. As one can see, the $\approx$20\,000 new RV stars
in the CRVAD-2 occupy an intermediate magnitude interval and do not
extend the CRVAD to a fainter magnitude limit.

The CRVAD-2 contains accurate equatorial coordinates J2000, proper motions 
and trigonometric parallaxes in the Hipparcos system, $BV$ magnitudes in 
the Johnson system, variability and multiplicity flags, and
component identifiers from the ASCC-2.5, 
as well as radial velocities from the source 
catalogues. Spectral types are given from two different data sets.
One spectral type information is taken either from the ASCC-2.5 (mostly
from Hipparcos and the PPM catalogue) or from the GCRV, where our preference
was given to the most detailed spectral type. In a second column, we list
the information from the Tycho-2 Spectral Type Catalog of Wright et al.~(2003) 
providing the most complete list of stars with very detailed spectral
classification. As in the CRVAD, the RVs are not always listed
with individual measuring errors $\epsilon_{RV}$. 
Only 71\% of the CRVAD-2 stars have $\epsilon_{RV}$ values copied from
the source catalogues. Further 21.5\% have RV quality indices 
($A, B, C, D, E,$ or $I$) taken from Dufolt et al (1995), which correspond to 
standard errors of 0.74, 1.78, 3.70, 7.40, 10.0 km/s or to the case of 
insufficient data, respectively. For the remaining 7.5\% of the CRVAD-2
stars $\epsilon_{RV}$ were not available. 

\begin{table}
\caption{Contents of the CRVAD-2 }
\label{tab_crvad}
\setlength{\tabcolsep}{2pt}
\begin{tabular}{rlcl}
\hline
& Label&Units&Explanations\\
\hline
1  &RAhour      &  h    &  Right Ascension J2000.0, epoch 1991.25\\
2  &DEdeg       & deg   &  Declination J2000.0, epoch 1991.25\\
3  &$e_{RA}$    & mas   & Standard error of Right Ascension\\
4  &$e_{DE}$    & mas   & Standard error of Declination\\
5  &Plx         & mas   &   Trigonometric parallax\\
6  &$e_{Plx}$   & mas   &  Standard error of parallax\\
7  &pmRA        & mas/yr & Proper Motion in RA$\cdot$cos(DE)\\
8  &pmDE        & mas/yr & Proper Motion in DE\\
9  &$e_{pmRA}$  & mas/yr &  Standard error of pmRA\\
10 &$e_{pmDE}$  & mas/yr &  Standard error of pmDE\\
11 &$B$         & mag   &  $B$ magnitude in Johnson system\\
12 &$V$         & mag   &  $V$ magnitude in Johnson system\\
13 &$e_B$       & mag   &  Standard error of $B$ magnitude\\
14 &$e_V$       & mag   &  Standard error of $V$ magnitude\\
15 &Scat        & mag   &  Scatter in magnitude from Hipparcos\\
   &            &       &  or Tycho-1\\
16 &v1          & ---   &  Known variability from Hipparcos\\
17 &v2          & ---   &  Variability from Tycho-1\\
18 &v3          & ---   &  Variability type from Hipparcos\\
19 &v4          & ---   &  Variability from CMC11\\
20 &d12         & ---   &  CCDM component identifier\\
21 &d3          & ---   &  Component identifier from Hipparcos\\
22 &d4          & ---   &  Duplicity from Tycho-1\\
23 &d5          & ---   &  Double/Multiple Systems flag from\\
   &            &       &  Hipparcos\\
24 &d6          & ---   &  Duplicity flag from PPM\\
25 &Sp$_1$      & ---   & Spectral type from ASCC-2.5 or GCRV\\
26 &Sp$_2$      & ---   & Spectral type from Wright et al.~(2003)\\
27 &HIP         & ---   &  Hipparcos number\\
28 &HD          & ---   &  HD number\\
29 &ASCC        & ---   &  ASCC-2.5 number\\
30 &I$_{sc}$    & ---   &  Index of source catalogue\\
31 &comp        & ---   &  Components of multiple star or \\
   &            &       &  duplicity flag\\
32 &$RV$        & km/s  &  Radial Velocity\\
33 &$e_{RV}$    & km/s  &  Mean standard error of the RV\\
34 &$q_{RV}$    & ---   &  Quality index of the RV (A,B,C,D,E,I)\\
35 &$N_{RV}$    & ---   &  Number of observations\\
36 &$N_{mult}$  & ---   &  Number of matches in the source\\
   &            &       &  catalogue for given entry\\
\hline
\end{tabular}
\end{table}

\section{Radial velocities of open clusters and stellar associations}

\subsection{Object list}

The most complete list of open clusters described by Dias et al.~(2002)
contains 1759 objects after the last update (v. 2.7, 2006/10/27).
This list does not contain all 650 objects for which 
Kharchenko et al.~(2005a, 2005b) determined a homogeneous 
set of parameters, since the latter
included seven cluster-like associations (Vel~OB2, Nor~OB5, Sco~OB4,
Sco~OB5, Sgr~OB7, Cyg~OB2, Cep~OB3). As there is no sharp distinction
between open clusters and associations in the literature, we decided to 
add all stellar associations from Mel'nik (2007) so that our 
extended list contained 1821 objects. Note that among the about 80
associations listed by Mel'nik (2007), there are many
objects which used to be named open clusters, and treated as such in 
Kharchenko et al.~(2005a) and Dias et al.~(2002).

\begin{figure}[t]
\includegraphics[width=80mm,height=80mm,angle=270]{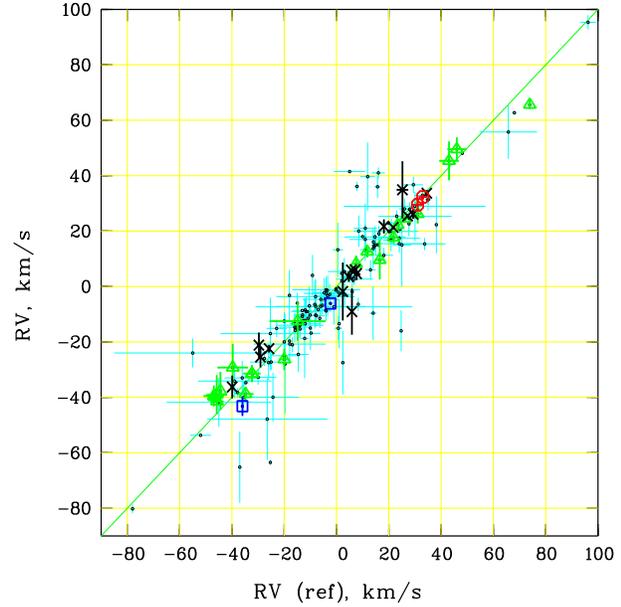}
\caption{Comparison of cluster $\overline{\rm{RV}}$s derived in this study
with published data (RV (ref)). Error bars indicate r.m.s. errors 
of the $\overline{\rm{RV}}$s (or of an individual RV measurement,
if only one cluster member star was available) 
Different symbols represent the comparison with the following sources:
black crosses -
Mermilliod et al.~(1998),
Mermilliod \& Mayor (1989),
Mermilliod \& Mayor (1990),
Mermilliod \& Mayor (1999),
Mermilliod, Mayor \& Burki (1987);
red circles -
Friel et al.~(2002),
Thorgersen, Friel \& Fallon (1993);
blue squares -
Turner (1992),
Turner, Forbes \& Pedreros (1992);
green triangles - 
Liu, Janes \& Bania (1991).}
\label{fig_comp_rv}
\end{figure}

\subsection{$\overline{\rm{RV}}$ determination on the basis of the CRVAD-2}

Based on the previously available RVs in the CRVAD, we had
obtained $\overline{\rm{RV}}$s of open clusters using their determined
members from the Catalogue of Stars in Open Cluster Areas (CSOCA) and its
first extension (Kharchenko at al. 2004b, 2005b). In this previous solution
(for a summary, see Scholz et al.~2005), 
we were able to get $\overline{\rm{RV}}$s for 322 out of 650 clusters
(here and in the following we call all 650 objects clusters, but keep
in mind that 7 of them are cluster-like associations)
identified in the ASCC-2.5. The data were presented together with other cluster 
parameters in the Catalogue of Open Cluster Data (COCD) and its first extension 
(Kharchenko at al. 2005a, 2005b).

Now, with the newly collected data in the CRVAD-2, we have updated the RV
solutions for a larger number of open clusters. A new sample of open cluster
members with RV measurements was obtained from a cross-identification of the
CRVAD-2 with cluster members from the CSOCA and its first extension.
The method for determining combined spatial-kinematical-photometric
membership probabilities $P_c$ has been described in detail by 
Kharchenko et al.~(2004b).

All possible members ($P_c > 1\%$) with RV measurements were involved in the 
determination of $\overline{\rm{RV}}$s. In all 650 clusters investigated there are 
about 38\,500 possible cluster members in total, from which only 4.8\% have
RV measurements. For 32 clusters, lacking such possible members with available
RVs, we used one bright RV star per cluster which could be considered as a member 
based on its proper motions only. 
This assumed cluster member lies on the upper main sequence or giant 
branch and probably failed our photometric membership
criterion due to intrinsic stellar variability and/or a non-uniform extinction
in the cluster area. In all these cases we re-analysed the location of the star 
in the proper motion diagram and in the colour-magnitude diagram, and the 
corresponding clusters are marked in the CRVOCA with $N_{\rm{RV}}=-1$. 

For 162 clusters there were more than two possible member stars with measured
RVs available (the average number of RV stars in these clusters was 9 with a 
maximum of 106). For each of these clusters we checked the RVs for consistency and
excluded stars with RVs deviating from the average by more than 3 standard 
deviations from our final solution. 
The average of the 162 standard errors of the mean cluster radial
velocities ($\overline{\rm{RV}}$s), determined from our selected
members and their RVs in the CRVAD-2, is 4.5~km/s.
For 60 clusters with two members and for 124 clusters
with only one member and available error of the individual stellar RV (formally
considered as the error of the cluster $\overline{\rm{RV}}$) the 
average 
errors
are 4.9 and 4.6 km/s, respectively. 17 clusters with $\overline{\rm{RV}}$s
obtained from the CRVAD-2 have no error estimate.

Altogether, we used 1749 stars from the CRVAD-2 for determining 
$\overline{\rm{RV}}$s of 363 out of 650 clusters. In case of 246 clusters
our previously determined $\overline{\rm{RV}}$s were confirmed, for 76
clusters we got improved values, and for 41 clusters we computed their
$\overline{\rm{RV}}$s for the first time based on our membership analysis
and available RVs in the CRVAD-2. 

\begin{figure}[t]
\includegraphics[width=80mm,height=80mm,angle=270]{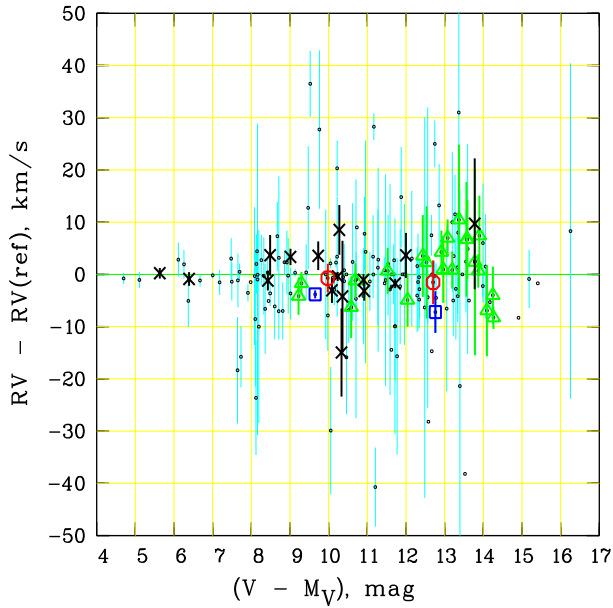}
\caption{Differences between published $\overline{\rm{RV}}$s and
$\overline{\rm{RV}}$s from this study 
versus cluster distance. The designations are the same
as in Fig.~\ref{fig_comp_rv}.}
\label{fig_drv06_dt}
\end{figure}

\subsection{Literature data}

Several major catalogues of $\overline{\rm{RV}}$s of open clusters and stellar
associations have been published so far:

Hron's (1987) $\overline{\rm{RV}}$s were calculated using some
weighting system for 105 young clusters with stars of earliest spectral types
(up to B3).  
Rastorguev et al.~(1999) published $\overline{\rm{RV}}$s of 117 open clusters
younger than 100 million years, where the data for 40 clusters were taken from 
Hron (1987). For the remaining 77 clusters they obtained $\overline{\rm{RV}}$s from
the WEBDA database (http://www.univie.ac.at/webda/), or from original observations 
by Glushkova \& Rastorguev (1991) with a correlation spectrometer.
Applying a cluster analysis technique to Galactic OB-stars Mel'nik \& 
Efremov (1995) identified 88 compact groups (association cores) and 
determined $\overline{\rm{RV}}$s for about half of them. 
A more complete list of associations with respect to $\overline{\rm{RV}}$ 
data was recommended to us by Mel'nik (2007). 
This list of 81 associations (73 of which have $\overline{\rm{RV}}$s)
is based on the partition of OB-stars
into associations suggested by Blaha \& Humphreys (1989).
It has been used by Mel'nik, Dambis \&
Rastorguev (2001) in a kinematical analysis of OB associations.
We prefered to use these data, available 
at http://lnfm1.sai.msu.ru/\~{}anna/page3.html, instead of those 
from Mel'nik \& Efremov (1995).

The catalogue of Lyng{\aa} (1987), not restricted to certain age classes,
contains $\overline{\rm{RV}}$s for 108 clusters (out of 1150 listed), for
which weighted mean values for member star velocities had been published
until 1980. The current Dias et al.~(2002) collection (v. 2.7, 2006/10/27)
of 1759 open clusters includes 361 $\overline{\rm{RV}}$s. 121 of the latter were
taken from our previous determinations based on the CRVAD.

Besides of the above mentioned catalogues 
there are several big projects which have contributed substantial numbers of
open cluster $\overline{\rm{RV}}$s
based on observations of cluster members selected by different methods:

One of the most productive projects by Mermilliod and co-workers
deals with observations of red giants and eclipsing binaries in open clusters.
In a large number of papers by
Clari\'a \& Mermilliod (1992),
Clari\'a, Mermilliod \& Piatti (1999),
Clari\'a et al.~(1994),
Clari\'a et al.~(2006),
Clari\'a et al.~(2003),
Eigenbrod et al (2004),
Huestamendia, del Rio \& Mermilliod (1990)
Mermilliod, Andersen \&Mayor (1997),
Mermilliod et al.~(1997),
Mermilliod et al.~(1995),
Mermilliod et al.~(2001),
Mermilliod et al.~(1996),
Mermilliod et al.~(2003),
Mermilliod et al.~(1998),
Mermilliod \& Mayor (1989),
Mermilliod \& Mayor (1990),
Mermilliod \& Mayor (1999),
Mermilliod, Mayor \& Burki (1987)
the $\overline{\rm{RV}}$s of 38 open clusters were determined.
In addition, Raboud \& Mermilliod (1998) and Robichon et al.~(1999)
determined $\overline{\rm{RV}}$s of 14 very nearby clusters based on
thoroughly selected cluster members.

\begin{figure}[t]
\includegraphics[width=80mm,height=80mm,angle=270]{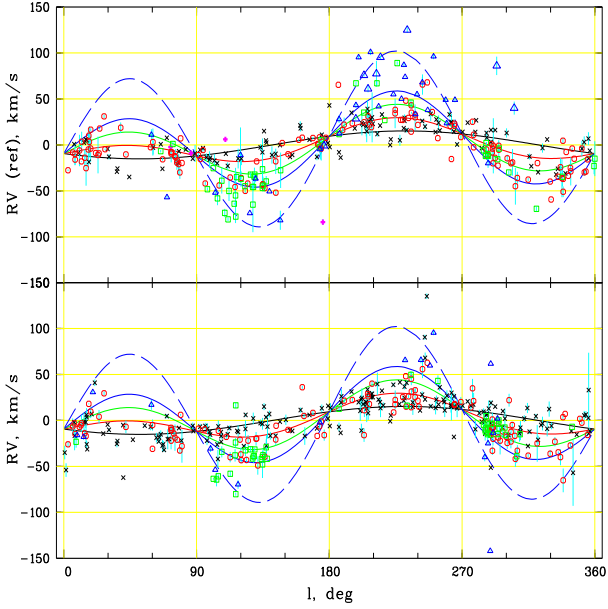}
\caption{Cluster $\overline{\rm{RV}}$s versus galactic longitude: determined
with CRVAD-2 (lower panel) and literature data (upper panel).
The curves show the combined effect of Solar motion and differential
Galactic rotation on the RVs with $d =$ 0, 1, 2, 3 and 6 kpc from the Sun
(black, red, green, blue, blue dashed lines, respectively).
Black crosses, red circles, green squares, blue triangles and large blue
triangles correspond to the clusters with $d =$ 0...1, 1...2, 2...3, 3...6,
and $>$6 kpc, respectively. Bold magenta pluses mark clusters with unknown $d$.}
\label{fig_rv_gl}
\end{figure}

\begin{figure}[t]
\includegraphics[width=45mm,height=80mm,angle=270]{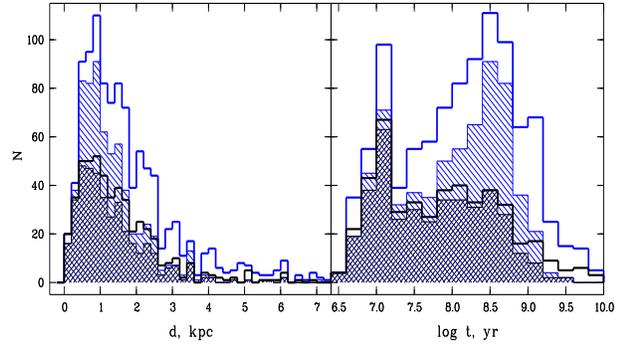}
\caption{Distributions of the open clusters with known $\overline{\rm{RV}}$s
and basic parameters with distance from the Sun (left panel) and with age
(right panel). Blue thick lines correspond to the objects with known
distances from the Sun (1047 objects) and known ages (997 objects), respectively.
Blue thin lines and hatched histograms show the sub-sample of
650 objects from Kharchenko et al.~(2005a, 2005b). Black thick lines
correspond to the objects with known $\overline{\rm{RV}}$s and distances
(512 objects), and with $\overline{\rm{RV}}$s and ages (465 objects), respectively.
Black thin lines and cross-hatched histograms represent 395 clusters from
Kharchenko et al.~(2005a, 2005b) with known $\overline{\rm{RV}}$s.  }
\label{fig_rv_his}
\end{figure}

An other big project carried out by
Friel (1989),
Friel \& Janes (1993),
Friel et al.~(2002),
Friel, Liu \&  Janes (1989),
Thorgersen, Friel \& Fallon (1993)
provided physical parameters, including $\overline{\rm{RV}}$s,
of 28 old open clusters.
Liu, Janes \& Bania (1991) measured $\overline{\rm{RV}}$s of
22 younger clusters.
High-quality membership selections of open clusters and accurate
$\overline{\rm{RV}}$s were also presented by
Turner (1992),
Turner, Forbes \& Pedreros (1992);
Turner (1993),
Turner, Mandushev \& Forbes (1994),
Turner, Pedreros \& Walker (1998).

All $\overline{\rm{RV}}$s, available in the above mentioned three catalogues,
in the WEBDA database and/or from the many publications
of original observational results on stellar RVs in open clusters and
associations, were critically reviewed
and added to our extended database of 1821 clusters and associations.
In the review of the literature RV data we tried to be as complete as possible
and to collect the most complete and homogeneous information concerning the
errors of $\overline{\rm{RV}}$s (some authors, e.g. Friel and co-authors,
list standard deviations instead of mean errors). 
The most reliable data were selected
according to more recent publications, to the largest number of cluster members,
and to the highest quality of the membership determination procedure.
In total, we found $\overline{\rm{RV}}$s
for 330 clusters and associations from 67 references in the literature.

215 out of the 330 clusters have published $\overline{\rm{RV}}$s based on at
least two (and up to 81) member stars. Their mean $\overline{\rm{RV}}$ error
is 4.1~km/s. For 25 clusters with only one member and for 66 clusters with an
unknown number of members the mean $\overline{\rm{RV}}$ errors from the 
literature are 2.6 and 6.2 km/s, respectively. 24 clusters have 
published $\overline{\rm{RV}}$s (based on one member or on an unknown 
number of members) without specified errors.
The latter are taken mainly from Lyng{\aa} (1987), Mel'nik (2007),
and Dias et al.~(2002).

\begin{table}
\caption{Contents of the CRVOCA}
\label{tab_crvoca}
\setlength{\tabcolsep}{2pt}
\begin{tabular}{rlcl}
\hline
& Label&Units&Explanations\\
\hline
1 &N$_{COCD}$ & ---    & Number in accordance with\\
  &           &        & Kharchenko et al.~(2005a, 2005b)\\
2 &Name       & ---    & NGC, IC or other common designation\\
3 &RAhour     &  h     & Cluster center in RA J2000.0 \\
4 &DEdeg      & deg    & Cluster center in Dec J2000.0\\
5 & r         & deg    & Angular radius of the cluster\\
6 &RV         & km/s   & $\overline{\rm{RV}}$ determined with CRVAD-2\\
7 &$e_{\rm{RV}}$& km/s &  Mean standard error of the $\overline{\rm{RV}}$\\
8 &$N_{\rm{RV}}$&---& Number of stars for $\overline{\rm{RV}}$ determination\\
9 &RV$_{ref}$     & km/s & $\overline{\rm{RV}_{ref}}$ from literature\\
10&$e_{\rm{RV}_{ref}}$&km/s& Mean standard error of the $\overline{\rm{RV}_{ref}}$\\
11&$N_{\rm{RV}_{ref}}$&---& Number of stars for $\overline{\rm{RV}_{ref}}$
                             determination\\
12& ref      & --- & $\overline{\rm{RV}_{ref}}$ source identifier\\
\hline
\end{tabular}\\
\rule{0mm}{2mm}
Note:
If several coordinates and angular radii were available, we selected
the data from one of the sources with the following order of priority:
Kharchenko et al.~(2005a, 2005b), Dias et al.~(2002), Mel'nik (2007).
\end{table}

\section{Statistics of $\overline{\rm{RV}}$s of open clusters and 
stellar associations}

Altogether there are 516 objects with known $\overline{\rm{RV}}$s in the
CRVOCA.
For 177 objects there are both $\overline{\rm{RV}}$s determined based on the
CRVAD-2 as well as previously published values available. Fig.~\ref{fig_comp_rv}
shows the comparison of the best measurements taken from the literature 
with our newly determined $\overline{\rm{RV}}$s. The clusters with
high-quality solutions in the literature are shown with bold symbols.

The mean difference for all 177 objects in common is 
$\overline{\rm{RV}_{ref} - \overline{\rm{RV}}} = 0.65 \pm 0.72$ km/s. 
From Fig.~\ref{fig_drv06_dt} one can see that the differences increase
with larger distances. This effect is not only a result of increasing
errors of the RV measurements with fainter stars, but also reflects 
the more problematic membership selection for distant clusters.
All absolute differences larger than 25~km/s are mainly caused
by a very small number of cluster members (and their uncertain
membership probabilities) in at least one of the determinations.
Low-accurate individual RV measurements did also play a role in
some cases.

We have analysed all $\overline{\rm{RV}}$s as a function of Galactic longitude
$l$ (Fig.~\ref{fig_rv_gl}). The curves with different amplitudes show the
expected change of the RVs over $l$ for different distances $d$ from the Sun,
if there is no intrinsic motion but just the combined effect of Solar motion and
differential Galactic rotation. For the computation of these curves we used
the Solar motion components and Oort constants as obtained from the sample 
of 650 clusters studied in Piskunov et al.~(2006). As one can see, all clusters,
with only few exceptions, have $\overline{\rm{RV}}$s typical of disk objects,
i.e. their deviations from the corresponding curves are smaller than about
50 km/s. 

As a result of our study, 516 of all currently known objects (28\% from 1821) 
and 395 objects in our sample of open clusters and stellar associations with
homogeneously determined parameters (61\% from 650) have $\overline{\rm{RV}}$
estimates. 
Fig.~\ref{fig_rv_his} illustrates the current status of the determination
of basic parameters of open clusters and associations (a similar illustration 
of the status achieved in our previous study based on the CRVAD can be found 
in Scholz et al.~2005).
The sample of 1821 objects is very inhomogeneous with respect to the
availability of basic parameters, e.g. only 57\% have distance estimates,
and only 55\% have ages determined. One should also keep in mind
that the parameters of the objects in 
the full sample of 1821 objects have been derived with different methods
and observations so that they can not be easily compared.
However, we can consider our sample
of 650 objects as a homogeneous sample, for which the
basic parameters 
(angular sizes, average proper motions and radial velocities,
reddening and distances, and ages)
have been obtained in a uniform way. As seen in the right 
part of Fig.~\ref{fig_rv_his}, the availability of $\overline{\rm{RV}}$s for the 
clusters in this sample is biased to younger clusters: 78\% of the clusters younger 
than $\log t=8.3$ have $\overline{\rm{RV}}$ values, whereas among older clusters 
only 37\% have $\overline{\rm{RV}}$s measured. Even for the nearby clusters, located 
within our completeness limit of 850 pc (Piskunov et al.~2006), there are
$\overline{\rm{RV}}$ measurements available for 65\% only.

\section{Summary and outlook}

Both catalogues, the CRVAD-2 with 54907 entries, and the CRVOCA of 516 Galactic
open clusters and stellar association, can be retrieved from the 
Centre de Donn\'ees astronomiques de Strasbourg (CDS), France
via ftp or http (ftp://cdsarc.u-strasbg.fr/pub/cats/, 
http://vizier.u-strasbg.fr).
A summary of the CRVAD-2 columns is shown in Table~\ref{tab_crvad}.
The contents of the CRVOCA  is shown in Table~\ref{tab_crvoca}.
More details describing the data in these two catalogues can be found in the
corresponding ReadMe files at the CDS.
These catalogues supersede the first CRVAD version (Paper~I)
and the previously listed RV data on open clusters 
in Kharchenko et al.~(2005a, 2005b). 
Note that mean proper motions of 650 open clusters can be found in  
Kharchenko et al.~(2005a, 2005b) and are not listed here again.

The CRVAD-2 represents our knowledge on RV measurements of relatively bright stars
distributed over the whole sky in the pre-RAVE era. Only about 2\% of the stars
in the ASCC-2.5, which includes the complete Tycho-2 catalogue and formed the
basis of the CRVAD-2, have RVs from previous measurements. The RAVE survey of the
southern sky (Steinmetz 2003) aims at RV measurements for a much larger number of
Tycho-2 stars. Whereas the CRVAD-2 does not extend the magnitude limit of the
first CRVAD version, the RAVE observing program includes large numbers of
fainter stars, too. The recently published first RAVE data release contains
25\,274 RV measurements of 24\,748 individual stars (Steinmetz et al.~2006).
About half of these first RAVE stars are ASCC-2.5 stars, but only a negligible 
number of less than 50 of them have RV data in the CRVAD-2. Already with the next 
RAVE data release, expected in early 2007, the number of new RV stars will 
probably approach the number of stars in the CRVAD-2. A next version of the 
CRVAD will contain the first and second RAVE data releases.

We have successfully used the CRVAD-2 data in the compilation of the
CRVOCA representing a big step forward in improving our knowledge on 
the $\overline{\rm{RV}}$s of Galactic open clusters and associations.
The first RAVE data release includes almost no data in the Galactic plane,
where most of the open clusters and associations are located. However, with
the upcoming next RAVE data releases we expect to get the results of some
dedicated observations of open cluster fields which we selected and proposed
on the basis of our sample of 650 objects. According to our proposal,
RAVE observations are currently under way or have already been finished
for more than 1500 members of about 100 open clusters in 12 selected fields
in the Galactic plane.
We expect that these observations will allow us to compute space motions 
of a large number of open clusters and to study, in more detail, the kinematics 
and membership of some individual clusters.

\acknowledgements
This work was supported by the DFG grant 436~RUS~113/757/0-2 and 
by the RFBR grant 06-02-16379.
We acknowledge the use of 
the Simbad database  and the VizieR Catalogue Service operated at the CDS,
the WEBDA facility at the Institute of Astronomy of the University of Vienna.
Finally, we would like to thank the referee, Dr. Allende Prieto, for his
comments/suggestions, which helped to improve the paper.

\newpage

\end{document}